\documentclass[12pt,a4paper]{iopart}
\usepackage{amssymb,graphicx,cite}

\usepackage{graphicx}
\begin{document}
   
%\twocolumn[\hsize\textwidth\columnwidth\hsize\csname@twocolumnfalse\endcsname
 
%\title[Jet formation in a collapsing Bose-Einstein condensate]{Mean-field
%model of jet formation in a collapsing Bose-Einstein condensate}

\title[Bright solitons in a
fermion-fermion mixture]{Formation of bright solitons and soliton 
trains in a 
fermion-fermion mixture by modulational instability}
 
\author{Sadhan K. Adhikari\footnote{Electronic
address: adhikari@ift.unesp.br; \\
URL: http://www.ift.unesp.br/users/adhikari/}}
%\affiliation
\address
{Instituto de F\'{\i}sica Te\'orica, UNESP $-$ S\~ao Paulo
State University,  01.405-900 S\~ao Paulo, S\~ao Paulo, Brazil}
 
\date{\today}
 
%\maketitle
 
\begin{abstract}

We employ a time-dependent  mean-field-hydrodynamic
model to  study the  generation of   bright solitons in a
degenerate
fermion-fermion mixture in a  cigar-shaped geometry
using variational and numerical methods.
Due to a strong Pauli-blocking repulsion among identical spin-polarized
fermions at short distances there cannot be bright  solitons for
repulsive interspecies interactions.  Employing a linear stability
analysis 
we demonstrate the formation of stable 
solitons due to modulational instability of a 
constant-amplitude solution
of the model equations  
for a sufficiently
attractive interspecies  interaction. 
We perform a 
numerical stability analysis of these solitons and also demonstrate the
formation of soliton trains  by jumping the effective interspecies
interaction from repulsive to attractive.  
These fermionic solitons can be formed and
studied in  laboratory with present technology.


\end{abstract}

\maketitle

%\keywords{Degenerate Fermi gas, Bright soliton, Soliton train}

%\date{\today}

%\maketitle

\section{Introduction} 

After  experimental observation and the study  of bright solitons in
Bose-Einstein condensates (BEC) \cite{exdks1,exdks2,yuka},  
recent observations  \cite{exp1,exp2,exp3,exp4}  and 
experimental \cite{exp5,exp5x,exp6} and theoretical
\cite{yyy1,yyy,capu,ska} studies
of  a degenerate Fermi gas (DFG) by
sympathetic cooling in
the presence of a second boson or fermion component suggest the
possibility of soliton formation \cite{vpg} in a degenerate
fermion-fermion     mixture (DFFM). 
Apart from the observation of a DFG in the 
degenerate boson-fermion
mixtures (DBFM)
 $^{6,7}$Li
\cite{exp3}, $^{23}$Na-$^6$Li \cite{exp4} and $^{87}$Rb-$^{40}$K
\cite{exp5,exp5x}, there have been studies of the following 
spin-polarized 
DFFM
 $^{40}$K-$^{40}$K \cite{exp1} and $^6$Li-$^6$Li \cite{exp2}.

  The dimensionless
one-di\-mensional
nonlinear Schr\"odinger (NLS)
equation in the attractive (self-focussing) case \cite{1}
\begin{equation}\label{nls} i \frac{\partial \phi}{\partial
t}+\frac{1}{2}\frac {\partial^2 \phi}{\partial y^2}+ |\phi|^2\phi =0.
\end{equation} sustains the following bright
 soliton \cite{1}: 
\begin{eqnarray}\label{DS}
\phi(y,t)&=&  a \hskip 3pt \mbox{sech}
[a(y-v t)] e^{ivy -i(v^2 -a^2)t/2+i\sigma}
\end{eqnarray}
 The parameter $a$ represents the amplitude as well as pulse width, $v$
the velocity, and $\sigma$ is a phase constant. These bright solitons 
are possible only in one dimension. In two and three dimensions they are 
allowed in the  presence of a transverse trap \cite{exdks1,vpg}, or 
in the presence of 
an 
oscillating nonlinearity \cite{soldy}.
Apart from  bright solitons in the attractive case, gap solitons are 
possible in the repulsive (self-defocussing) case  in the presence of a 
periodic potential \cite{gs}.

Bright solitons in a 
BEC are formed due to a nonlinear atomic
attraction \cite{exdks1,exdks2}.  As the interaction in a pure DFG at
short
distances is
repulsive due to strong Pauli blocking, there cannot be bright solitons in
a DFG. 
However, bright
solitons
can be
formed \cite{fbs1,fbs2}
in a DBFM in the presence of a sufficiently strong
boson-fermion attraction which can overcome the Pauli repulsion among
identical fermions. Bright solitons can also be formed in a binary
mixture of repulsive bosons supported by interspecies attraction
\cite{perez}.

We demonstrate the formation of stable fermionic bright solitons in a DFFM
for a sufficiently attractive interspecies interaction.
In a DFFM, the coupled system can lower its energy by forming high density
regions, the bright solitons, when the attraction between the two types of
fermions is large enough to overcome the Pauli repulsion. We use a coupled time-dependent mean-field-hydrodynamic model
for a DFFM and consider the formation of axially-free localized bright
solitons in a quasi-one-dimensional cigar-shaped geometry using numerical
and variational solutions.  The present model is inspired by the success
of a similar model used recently in the investigation of collapse
\cite{ska,skal} and bright \cite{fbs2} and dark \cite{fds} solitons in a
DBFM, and black solitons \cite{bla}
and mixing-demixing \cite{mix} in a DFFM.

We study the condition of modulational instability of a constant-amplitude
solution of the present model under a plane-wave perturbation and
demonstrate the possibility of the formation of bright solitons by a
linear stability analysis.  We find that for a sufficiently strong
interspecies interaction, under the plane-wave perturbation the
constant-amplitude solution becomes unstable and localized solitonic
solutions may appear.  We present a numerical stability analysis of these
bright solitons by introducing different small perturbations when the
solitons undergo stable and sustained breathing oscillation.  

We also
consider and study the formation of a soliton train in a DFFM by a large
sudden jump in the interspecies fermion-fermion scattering length 
realized by manipulating a background magnetic field near a Feshbach 
resonance,  experimentally observed in two hyperfine 
states of the
following spin-polarized 
DFFMs: $^6$Li-$^6$Li and $^{40}$K-$^{40}$K \cite{fesh}.
This procedure effectively and rapidly turns a repulsive DFFM to a 
highly attractive one and generates bright solitons. 
An experimental realization of bright solitons in a DFFM could be tried
in two hyperfine states in samples such as $^6$Li-$^6$Li or
$^{40}$K-$^{40}$K using a Feshbach resonance.  
However, there is already experimental evidence \cite{gehm} and 
theoretical 
conjecture \cite{kkk}
that the maximum attractive force, that can be created by a 
Feshbach resonance in fermionic atoms in two hyperfine states is 
limited 
by quantum mechanical constraints of unitarity.  Although  there is a 
limit to the creation of attraction in  a two-component spin-polarized 
DFFM  in two hyperfine states, there is no such limit in a 
multi-component 
spin-polarized
DFFM \cite{hei}
or a DFFM composed of atoms of distinct mass \cite{hulet1}.
So, if bright solitons and soliton trains cannot be efficiently created 
in a  two-component DFFM  in two hyperfine states,
a better and more efficient  DFFM for 
the creation of solitons could be the mixture of  
fermionic atoms of distinct mass, where one can avoid the problem of a 
possible suppression of interspecies attraction. One such system, among 
many others,  could 
be the mixture of spin-polarized $^6$Li-$^{40}$K mixture: both $^6$Li 
\cite{exp2,exp211} 
and $^{40}$K \cite{exp1} have 
been trapped and studied in laboratory.
The formation of bright solitons in a DFFM by turning the effective 
interspecies interaction among spin-polarized
fermions from repulsion to strong attraction 
seems within the reach of present experimental possibilities.

Here we shall be interested in the formation of bright solitons in a
spin-polarized DFFM of different fermionic atoms in the presence of
strong interspices attraction and we shall ignore the possibility of the
formation of a Bardeen-Cooper-Schreiffer (BCS) condensate \cite{bcs}
through Cooper pairing. A BCS condensate is usually formed in the
presence of a weak attraction among identical fermions with fermion
pairs in the singlet (spin parallel) state. The formation of a BCS
condensate should not be favored in a spin-polarized (spin antiparallel)
two-component DFFM with strong attraction among the different types of
atoms, as a strong interspecies attraction is not the domain of BCS
condensation. By choosing a strong interspecies attraction our study
stays in the BEC region of the BCS-Bose crossover problem \cite{BCSB}
strongly favoring molecule formation. The solitons once formed will
decay via molecule formation of two different types of fermionic atoms
\cite{hulet3} as in the case of bosonic solitons.  Nevertheless, such
molecule formation is a slow process and can be accounted for by a
three-body recombination term as introduced in a previous study of
collapse in a DFFM \cite{njp}. The molecule formation will eventually
destroy the solitons in the DFFM, albeit, at a slow rate. The same is
true in the formation of soliton and soliton train in an attractive BEC,
where the essential features of the dynamics have been well explained
within the mean-field Gross-Pitaevskii model by neglecting molecule
formation \cite{hulet2}. A subsequent study including the effect of
molecule formation \cite{sala2} has not essentially changed the
conclusions of \cite{hulet2}.  Hence this pioneering study on the
formation of soliton and soliton trains in a DFFM of different atoms
using mean-field hydrodynamic equations with the neglect of molecule
formation is expected to explain the essential features of the dynamics. 
However, it would be of interest to include the effect of molecule 
formation on the DFFM solitons in a future study.

In section 2 we present our mean-field-hydrodynamic model and its 
reduction
to a quasi-one-dimensional form in a cigar-shaped geometry.  In section 
3
we show that bright solitons can appear in this model
through modulational instability of a constant amplitude solution. In 
section 4 we perform a variational analysis of the mean-field equations.
Numerical results for isolated bright solitons are presented in section 
5 and
are compared with variational results. Then we study the generation of 
a
train of solitons by modulational instability. Finally, we present a
summary in section 6.

 \section{Nonlinear Model}

We  use a  simplified mean-field-hydrodynamic Lagrangian for 
a DFG used successfully to study a DBFM
\cite{ska,fbs2,fds}. 
%The virtue of the
%mean-field model over a  microscopic description is its simplicity and
%predictive power. 
To  develop a set of  time-dependent
mean-field-hydrodynamic
equations for the interacting DFFM, we use   the
following Lagrangian density \cite{ska,fbs2} 
\begin{eqnarray}\label{yy} {\cal
L}= g_{12}n_1n_2+\sum_{j=1}^2
\frac{i}{2}\hbar \left[ \psi_j\frac{\partial {\psi_j} ^*}{\partial
t} - {\psi_j}^* \frac{\partial \psi_j}{\partial t} \right]+
\end{eqnarray}    \begin{eqnarray}
+ \sum_{j=1}^2     
\left(\frac{\hbar^2 |\nabla_{\bf r} \psi_j|^2 }{6m_j}+
V_j({\bf r})n_j+\frac{3}{5} A_j n_j^{5/3}\right),\nonumber
\end{eqnarray} 
where $j=1,2$ represents the two components, $\psi_j$ the  probability 
amplitude, $n_j=|\psi_j| ^2$ the probability 
density,  
$^*$ denotes complex conjugate,  $m_j$  the
mass,   
$A_j=\hbar^2(6\pi^2)^{2/3}/(2m_i),$ 
the interspecies coupling        
$g_{12}=2\pi \hbar^2 a_{12} 
/m_R$ 
with $m_R=m_1m_2/(m_1+m_2)$ the reduced mass,  and 
$ a_{12}$ 
 the interspecies 
fermion-fermion scattering length.
 The
number of fermionic atoms $N_j$
is given by  $\int d{\bf r} n_j({\bf r})=N_j$.
The trap potential with axial symmetry is  
taken as $
V_{j}({\bf
r})=\frac{1}{2} 3m_j \omega ^2 (\rho^2+\nu^2 z^2)$ where
 $\omega$ and $\nu \omega$ are the angular frequencies in the radial
($\rho$) and axial ($z$) directions with $\nu$ the anisotropy.
The  $\nu \to 0$ limit corresponds to a cigar-shaped geometry and 
allows a
reduction of the three-dimensional equations to a quasi-one-dimensional
form appropriate for freely moving solitons. 
 The interaction between identical intra-species fermions in
spin-polarized state is highly suppressed due 
to Pauli blocking terms $3A_jn_j^{5/3}/5$ 
and has been neglected in   (\ref{yy}).   The
kinetic energy terms $\hbar^2|\nabla_{\bf r}\psi_j|^2$
$/(6m_j)$
in   (\ref{yy})
contribute little to this problem compared to the
dominating Pauli-blocking terms.  
However, its inclusion leads
to an analytic solution for the probability density everywhere
\cite{fbs2}.

With the Lagrangian density (\ref{yy}), the following Euler-Lagrange
equations: 
\begin{equation}
\frac{d}{d t}\frac{\partial {\cal L}}{\partial
\frac{\partial
\psi_j\- ^*}{\partial t}}+
\sum _{k=1}^3 \frac{d}{dx_k}\frac{\partial {\cal L}}{\partial
\frac{\partial \psi_j\- ^*}{\partial x_k}}= \frac{\partial {\cal
L}}{\partial
 \psi_j\- ^*},
\end{equation}
with $x_k, k=1,2,3$ being the three space components, 
become
\begin{eqnarray}\label{e} \biggr[ 
i\hbar\frac{\partial }{\partial t} +\frac{\hbar^2\nabla_{\bf
r}^2}{6m_{{j}}} - V_{{j}} - A_jn_j^{2/3}-
g_{{12}}
n_k
 \biggr]\psi_j=0,
\end{eqnarray}
where $j\ne k = 1,2$. This is a time-dependent version of a
similar time-independent hydrodynamic 
model for fermions \cite{capu}.
For large $n_j$, both lead to \cite{capu,ska}
the
Thomas-Fermi result \cite{yuka} $n_j=[(\mu_j-V_j)/A_j]^{3/2}$
with $\mu_j$ the chemical potential.   They yield  identical  results
for  time-independent stationary states.
However, 
the
present time-dependent 
model  can be used in the study of
nonequilibrium dynamics, as in the study of soliton trains.

For  $\nu =0$,   (\ref{e})
 can be reduced to an effective
one-dimensional form by considering
solutions of the type
$\psi_j({\bf r},t)=\sqrt N_j  \phi_j(z,t)\psi_j^{(0)}( \rho)$ 
where
\begin{eqnarray}\label{wfx}
|\psi_j^{(0)}(\rho)|^2&\equiv&
{\frac{m\omega}{\pi\hbar}}\exp\left(-\frac{m
\omega
\rho^2}{\hbar}\right),
\end{eqnarray}
with 
$m=3m_j $ 
corresponds to the 
ground state wave function in the radial trap alone 
in the absence of 
nonlinear interactions. Here to have an algebraic simplification we have
taken the masses of two types of fermions to be equal ($m_1=m_2$).
These wave functions satisfy
\begin{eqnarray}
-\frac{\hbar^2}{2m}\nabla_\rho ^2\psi_j^{(0)}
+
\frac{1}{2}m\omega^2\rho^2
\psi_j^{(0)}&=&\hbar\omega
\psi_j^{(0)},
\end{eqnarray}
with normalization
$2\pi \int_{0}^\infty |\psi_j^{(0)}(\rho)|^2 \rho d\rho=1.$
Now the dynamics is carried by $ \phi_j(z,t)$ and the radial dependence is
frozen in the ground state $\psi_j^{(0)}(\rho)$.
In the quasi-one-dimensional cigar-shaped
geometry the
linear fermionic probability densities are given by 
$|\phi_j(z,t)|^2$.

Averaging over the radial mode $\psi_i^{(0)}(\rho)$,
i.e., multiplying
 (\ref{e}) 
by  $\psi_i^{(0)*}(\rho)$
and integrating over $\rho$, we obtain the following one-dimensional
dynamical equations \cite{mix}:
\begin{eqnarray}\label{i}
\biggr[  i \hbar\frac{\partial
}{\partial t}
+\frac{\hbar^2}{2m}\frac{\partial^2}{\partial z^2}
%\nonumber \\
-F_{jj}|
\phi_j|^{4/3}
- F_{jk}| \phi_k|^2
 \biggr] \phi_j(z,t)=0, \nonumber \\
\end{eqnarray}
where
\begin{eqnarray}
 F_{jk}=g_{12}N_k\frac{\int_0^\infty|\psi_j^{(0)}|^2|\psi_k^{(0)}|^2
\rho d\rho}
{\int_0^\infty|\psi_j^{(0)}|^2\rho d\rho}=
g_{12}{\frac{N_km\omega}{2\pi\hbar}}, \nonumber
\end{eqnarray} 
\begin{eqnarray}
F_{jj}=A_jN_j^{2/3}
\frac{\int_0^\infty|\psi_j^{(0)}|^{2+4/3}\rho
d\rho}{\int_0^\infty|\psi_j^{(0)}|^2\rho
d\rho} =
{\frac{3A_j}{5}}\left[
\frac{N_jm\omega}{\pi \hbar}    \right]^{2/3}.\nonumber 
\end{eqnarray}
In   (\ref{i}) 
the normalization
is given by $\int_{-\infty}^\infty |\phi_j(z,t)|^2
dz = 1$. In these equations we have set the anisotropy parameter
$\nu=0$
 to remove the axial trap and thus to generate axially-free
quasi-one-dimensional solitons.

To reduce  three-dimensional equations  (\ref{i})  
to a dimensionless form,
following  \cite{fbs2},  
  we consider variables
$\tau=t \omega/2$, $y=z /l$, ${\varphi}_i= \sqrt l \phi_i$, with
$l=\sqrt{\hbar/( \omega m)}$, while   (\ref{i}) becomes 
\begin{eqnarray}\label{m} \biggr[ i\frac{\partial
}{\partial \tau}
+\frac{\partial ^2}{\partial y^2} 
-    
 N_{jj}
\left|{{\varphi}_j}\right|^{4/3}        
+N_{jk}
  \left|{{\varphi}_k}\right|^2                  
 \biggr]{\varphi}_{{j}}({y},\tau)=0,         
\nonumber \\
\end{eqnarray}
where   
$N_{jj}=9(6\pi N_j)^{2/3}/5, $ and 
$N_{jk}=12 |a_{12}|N_k/l$. Here we employ  
a negative $a_{12}$ corresponding to
attraction, and     
$\int_{-\infty}^\infty |\varphi_j(y,\tau)|^2 dy =1 .$
In   (\ref{m})  
a sufficiently strong  
attractive  fermion-fermion coupling
 $N_{jk}|\varphi_k|^2 (j\ne k)$ can  overcome  the Pauli 
repulsion $N_{jj}|\varphi_j|^{4/3}$
and  form  bright solitons.

For the usual Gross-Pitaevskii with the cubic nonlinearity, the 
reduction of the full 
three-dimensional
equation to its one-dimensional counterpart was performed, in different 
forms, by 
many
authors \cite{Luca}.
The nonlinearity with 4/3 power in (\ref{m}) was obtained starting from 
a similar nonlinearity in a three-dimensional DFG. A strict 
one-dimensional (two-dimensional) DFG will give rise to  a quintic 
(cubic)
nonlinearity. Whatever be the nonlinearity in the intraspecies fermions, 
bright fermionic solitons will be generated, provided that the 
interspecies attraction is strong enough to overcome the intraspecies 
repulsion due to Pauli blocking. However, in this paper we shall 
consider   only the Pauli-blocking nonlinearity with the 4/3 power as in 
(\ref{m}).

The two coupled equations (\ref{m}) could be simplified for $N_1=N_2=N$,
while  
$\varphi_1=\varphi_2 \equiv \varphi$, and 
these equations reduce to the following single
equation
\begin{eqnarray}\label{o} \biggr[ i\frac{\partial
}{\partial \tau}
+ \frac{\partial ^2}{\partial y^2}
-    \beta
\left|{\varphi}\right|^{4/3}
+\gamma
  \left|{{\varphi}}\right|^2
 \biggr]{\varphi}({y},\tau)=0,
\end{eqnarray}
where $\beta=N_{11}=N_{22}$ and $\gamma=N_{12}=N_{21}$. Equation
(\ref{o}) maintains the essential features of   (\ref{m}), e.g., a
quadratic nonlinear attraction and a Pauli blocking repulsion.
The one-dimensional 
equation  (\ref{o}) is quite similar in structure to the NLS equation
(\ref{nls}) apart for the Pauli-blocking repulsive term  $\beta$.
For a small $\beta$ and large
$\gamma$ 
the solitons of  
 (\ref{nls}) survive in   (\ref{o}). However, they disappear in the
opposite limit of large  $\beta$ and small  $\gamma$.

\section{Modulational Instability}

\subsection{Symmetric Case ($N_1=N_2$)}

We find that 
   (\ref{o}) allows a constant-amplitude 
        solution which 
exhibits modulational instability leading  
to a modulation of
the          solution. 
We perform a
stability analysis  of this  
solution 
 and  study the
possibility of generation 
of  solitons by modulational instability 
in the symmetric case. 
We consider the 
constant-amplitude solution \cite{1} 
\begin{equation}\label{so}
\varphi_0=A_0
\exp(i\delta) \equiv A_0 \exp [i(\gamma A_0^2 \tau -\beta A_0^{4/3}
\tau)]
\end{equation}
 of  (\ref{o}),
where  $A_0$ is the constant amplitude and
$\delta$ a phase. 
The time evolution of  solution (\ref{so}) 
maintains the constant amplitude $A_0$ but acquires an amplitude-dependent
phase.  Now we study if this 
 solution is stable
against small perturbations by performing a linear stability analysis. 

We consider  a small perturbation of the constant-amplitude 
solution
(\ref{so}) given by: 
\begin{equation}\label{per}
\varphi=(A_0+ A)\exp(i \delta),
\end{equation}
where  $A=A(y, \tau)$  is the 
small perturbation.
Substituting the
perturbed solution  (\ref{per})
in   (\ref{o}), and for small perturbations retaining
only the linear terms in $A$ we get 
\begin{eqnarray}\label{p} i\frac{\partial A
}{\partial \tau}
+ \frac{\partial ^2 A}{\partial y^2} 
-   \frac{2}{3} \beta A_0^{4/3}(A+A^*)
+\gamma A_0^2 (A+A^*)=0.\nonumber \\
\end{eqnarray}

We consider the complex plane-wave
perturbation  
\begin{equation}
A(y,\tau)=
{\cal A}_1 \cos (K\tau -\Omega y)+i {\cal A}_2 \sin  (K\tau -\Omega y)
\end{equation} 
in  
(\ref{p}), where ${\cal A}_1$ and ${\cal A}_2$ are the amplitudes of the
real and imaginary parts, respectively, and 
$K$ is a frequency parameter and 
$\Omega$ a wave number. Then separating the
real and imaginary parts we get
\begin{eqnarray}
-{\cal A}_1 K & =& {\cal A}_2 \Omega^2,\\
-{\cal A}_2K &=& {\cal A}_1 \Omega^2   -2\gamma A_0^2
{\cal A}_1+\frac{4}{3}\beta A_0^{4/3}{\cal A}_1,
\end{eqnarray}
and 
eliminating  ${\cal A}_1$ and ${\cal A}_2$ we obtain the dispersion
relation 
\begin{equation}\label{di}
K=\pm  \Omega \left[\Omega^2-({2\gamma A_0^2-\frac{4}{3}\beta
A_0^{4/3}})\right]^{1/2}.
\end{equation}
The constant-amplitude solution (\ref{so}) is stable if perturbations 
at any wave number  $\Omega$ do not grow with time. This is true as long
as frequency $K$ is real. From   (\ref{di}) we find that 
 $K$ remains real for any $\Omega$   provided that 
$2\gamma A_0^2 < 4\beta A_0^{4/3}/3$ or $\gamma A_0^{2/3}
< 2\beta /3$.  However, $K$ can become imaginary for
$\gamma A_0^{2/3}
> 2\beta /3$  and the plane-wave perturbations can grow exponentially
with time $\tau$. This is the domain of modulational instability 
of a constant-intensity solution \cite{1}. The perturbation then grows 
exponentially with the intensity given by the growth rate or the
modulational instability gain $g(\Omega)$
defined by 
\begin{equation}
g(\Omega)\equiv  2 \Im (K)
= 2|\Omega|\biggr[ 2\gamma A_0^2-\frac{4}{3}\beta
A_0^{4/3}-\Omega^2 \biggr]^{1/2},
\end{equation}
where $\Im$ denotes imaginary part.
The  presence of modulational instability  is closely
connected
with the appearance of a bright soliton \cite{1}. 
Localized bright
solitons are possible only when the constant-amplitude solution is
unstable.

\subsection{Asymmetric Case ($N_1\ne N_2$)}

Now we consider the possibility of modulational instability of a similar 
constant-intensity solution in coupled 
equations (\ref{m}). We consider the 
constant-amplitude solutions 
\begin{eqnarray}
\varphi_{j0}=A_{j0}\exp(i\delta_j)\equiv A_{j0}\exp[i\tau(N_{jk}
A_{j0}^2
-N_{jj}  A_{j0}^{4/3} )], \nonumber
\end{eqnarray}
of   (\ref{m}), where $A_{j0}$ is the amplitude and $\delta_j$ a phase
for component $j$. The constant-amplitude solution develops an
amplitude dependent phase on time evolution. 
We consider a  small perturbation $A_j\exp(i\delta_j)$ to 
these solutions via  
\begin{eqnarray} 
\varphi_j=(A_{j0}+A_j)\exp(i\delta_j),
  \end{eqnarray}              
where $A_j=A_j(y,\tau)$. 
Substituting this perturbed solution in   (\ref{m}), 
and for small perturbations retaining
only the linear terms in $A_j$ we get
\begin{eqnarray}\label{q} 
i\frac{\partial A_j
}{\partial \tau}
+ \frac{\partial ^2 A_j}{\partial y^2}
-   \frac{2}{3} N_{jj} A_{j0}^{4/3}(A_j+A_j^*)+
\end{eqnarray}\begin{eqnarray}
+N_{jk}
A_{k0}A_{j0} (A_k+A_k^*)=0, \quad j\ne k.\nonumber
\end{eqnarray}
We consider the complex plane-wave
perturbation  
\begin{equation}
A_j(y,\tau)=
{\cal A}_{j1} \cos (K\tau -\Omega y)+i {\cal A}_{j2} \sin  (K\tau -\Omega
y)\nonumber
\end{equation} 
in  
 (\ref{q}) for $j=1,2$, where ${\cal A}_{j1}$ and  ${\cal A}_{j2}$ 
are the amplitudes for the real and imaginary parts, respectively, and 
$K$ and $\Omega$ are frequency and wave numbers. 
Then separating the real and imaginary parts we
get 
\begin{eqnarray}\label{p1}
-{\cal A}_{11}K&=&{\cal A}_{12}\Omega^2 \\
-{\cal 
A}_{12}K&=&{\cal 
A}_{11}\Omega^2-2N_{12}A_{10}A_{20}{\cal A}_{21}+\frac{4}{3}N_{11}
A_{10}^{4/3}{\cal A}_{11}, \nonumber \\ \label{p2} 
  \end{eqnarray}
for  $j=1$, and 
\begin{eqnarray}\label{p3}    
-{\cal   A}_{21}K&=&{\cal   A}_{22}\Omega^2 \\
-{\cal 
A}_{22}K&=&{\cal 
  A}_{21}\Omega^2-2N_{21}A_{10}A_{20}{\cal A}_{11}+\frac{4}{3}N_{22}
A_{20}^{4/3}{\cal   A}_{21},   \nonumber \\  \label{p4}            
  \end{eqnarray}
for $j=2$. 
Eliminating ${\cal   A}_{12}$ between   (\ref{p1})  and
(\ref{p2}) we
obtain
\begin{eqnarray}\label{p5}
{\cal   A}_{11}[K^2-\Omega^2( \Omega^2+4N_{11}   A_{10}^{4/3}]=
2{\cal A}_{21}N_{12}A_{10}A_{20} \Omega^2,
\nonumber \\  \end{eqnarray}       
and eliminating ${\cal A}_{22}$ between  (\ref{p3})  and
(\ref{p4}) we
obtain
\begin{eqnarray}\label{p6}
{\cal A}_{21}[K^2-\Omega^2( \Omega^2+4N_{22}A_{20}^{4/3}]=
2{\cal A}_{11}N_{21}A_{10}A_{20} \Omega^2.
\nonumber \\  \end{eqnarray}
Eliminating ${\cal A}_{11}$ and ${\cal A}_{21}$ 
from   (\ref{p5}) and (\ref{p6}), 
finally,   we obtain the following
dispersion relation 
\begin{eqnarray}\label{p7}
K^2= \pm \Omega\biggr[ \biggr(\Omega^2+ 
\frac{2}{3}N_{11}A_{10}^{4/3}
+
\frac{2}{3}N_{22}A_{20}^{4/3}\biggr) \pm
\end{eqnarray}\begin{eqnarray}
\pm \biggr\{\frac{4}{9} \left( 
N_{11}A_{10}^{4/3} -
N_{22}A_{20}^{4/3} \right)^2
+
4N_{12}N_{21}
A_{10}^2A_{20}^2 \biggr\}^{1/2} \biggr]^{1/2}.\nonumber 
    \end{eqnarray}

For stability of the plane-wave perturbation,  $K$ has to be
real. For any
$\Omega$ this happens for 
\begin{eqnarray}
\biggr(N_{11}A_{10}^{4/3} 
+N_{22}A_{20}^{4/3}\biggr)^2  \nonumber 
\end{eqnarray}  \begin{eqnarray}  
> \left(
N_{11}A_{10}^{4/3} -
N_{22}A_{20}^{4/3} \right)^2+
9N_{12}N_{21}
A_{10}^2A_{20}^2, \nonumber 
\end{eqnarray}
or for $N_{12}N_{21}A_{10}^{2/3}A_{20}
^{2/3}< 4 N_{11}N_{22}/9.$ 
However, for 
$\-N_{12}N_{21}A_{10}^{2/3}A_{20}^{2/3}> 
4 N_{11}N_{22}/9$
\cite{shuk}, $K$ can become imaginary and the plane-wave perturbation can
grow exponentially with time. This is the domain of modulational
instability of a constant-intensity solution signalling the possibility
of coupled fermionic bright soliton to appear. In the symmetric case
$N_1=N_2$ and $A_{10}=A_{20}=A_0$, consequently,  $N_{11}=N_{22}=\beta$
and  $N_{12}=N_{21}=\gamma$ 
and we recover the condition of modulational instability 
$\gamma A_0^{2/3}> 2 \beta/3$ derived in section 3.1.

\section{Variational Analysis}

\subsection{Symmetric Case ($N_1=N_2$)}

Next we present a 
variational analysis of   (\ref{o}) based on the
Gaussian trial wave function \cite{and}
\begin{equation}\label{v}
\varphi_v(y,\tau)=A\exp\biggr[ -\frac{y^2}{2R^2(\tau)}+\frac{i}{2}
b(\tau)y^2+ic(\tau) \biggr], 
\end{equation}
where $A$ is the amplitude,  $R$ is the width,   $b$ the chirp, and $c$
the phase. 
The Lagrangian density for   (\ref{o})
is the one-term version of   
(\ref{yy}), e.g., 
\begin{eqnarray}
{\cal L}=\frac{i}{2}\hbar \biggr[\varphi
\frac{\partial \varphi^*
}{\partial t } - \varphi ^* \frac{\partial \varphi
}{\partial t }
\biggr]+ \biggr|\frac{\partial  \varphi}{\partial y} \biggr|^2
-\frac{1}{2}
\gamma n ^2 
+ \frac{3}{5}\beta n^{
5/3},\nonumber 
\end{eqnarray}
 which is evaluated with this 
trial function and the effective
Lagrangian  $L
=\int_{-\infty}^\infty {\cal L}(\varphi_v)dy$ becomes 
\begin{eqnarray}
L =\frac{A^2R\sqrt \pi}{2}\biggr(
\frac{6\sqrt 3}{5\sqrt 5}
\beta  A  ^{4/3}+2\dot c  -
\frac{\gamma }{\sqrt 2}  A ^2          
+
\end{eqnarray}
\begin{eqnarray}   
+\frac{R^2\dot b}{2} +
\frac{1}{R^2}
+b^2R^2
\biggr),\nonumber 
\end{eqnarray}
where the overhead dot denotes time derivative.
The  variational Lagrange
equations   %\cite{gold}
\begin{equation}\label{la}
\frac{d}{dt}\frac{\partial L}{\partial \dot q}= \frac{\partial
L}{\partial q},
\end{equation}
where $q$ stands 
for $c$, $A$,   $R$ and $b$
can then be written as 
\begin{eqnarray}\label{con}
A^2 R = \frac{1}{\sqrt \pi}=\mbox{constant}.
\end{eqnarray}
\begin{eqnarray}
\dot b&=& -\frac{2}{R^4}-2b^2+2\sqrt 2\gamma \frac{A^2}{R^2}
-4\frac{\sqrt 3}{\sqrt 5}\beta \frac{A^{4/3}}{R^2}-\frac{4\dot
c}{R^2},\label{x}\\
3\dot b&=& \frac{2}{R^4}-6b^2+\sqrt 2 \gamma \frac{A^2}{R^2}
-\frac{12}{5}\frac{\sqrt 3}{\sqrt 5}\beta \frac{A
^{4/3}}{R^2}-\frac{4\dot c}{R^2},\label{y}\\
\dot R &=&  2 R b.\label{z}
\end{eqnarray}
The constant in   (\ref{con}) is fixed by the normalization condition. 
Eliminating $\dot c$ from 
  (\ref{x}) and (\ref{y}) we obtain 
\begin{eqnarray} \label{eli}
\dot b = \frac{2}{R^4}-2b^2-\frac{1}{\sqrt 2} \gamma\frac{A ^2}{R^2}
+\frac{4}{5}\frac{\sqrt 3}{\sqrt 5} \beta \frac{A^{4/3}}{R^2}. 
 \end{eqnarray}      The use of 
  (\ref{con}), (\ref{z}) and (\ref{eli}) 
 leads to the following  
differential equation for the width $R$:
\begin{eqnarray} \label{min}
\frac{d^2 R}{d\tau^2}& =& 
\biggr(
\frac{4}{R^3}- \frac{\gamma}{R^2} \sqrt\frac{2}{\pi}
+\frac{1}{R^{5/3}}\frac{8\beta \sqrt
3}{5\pi^{1/3}
\sqrt5} \biggr)\\
&=&-\frac{d}{dR}\biggr[  \frac{2}{R^2}-\frac{\gamma}{R}
\sqrt\frac{2}{\pi}      +
\frac{1}{R^{2/3}}\frac{12\beta \sqrt
3}{5\pi^{1/3}
\sqrt5}
\biggr]. 
\end{eqnarray}
The quantity in the square bracket is the effective potential of the
equation of motion. Small oscillation around a stable configuration is
possible when there is a minimum in this potential. 
 The variational result for width $R$ follows by setting the
right hand side of   
(\ref{min}) to zero corresponding to a minimum in
this effective potential,  from which the variational 
profile for the soliton can be obtained \cite{and}. 

\subsection{Asymmetric Case ($N_1\ne N_2$)}

The above variational analysis can be extended to the asymmetric
case. However, the algebra becomes quite involved if we take a general
variational trial wave function with  chirp and phase parameters. As we
are interested mostly in the density profiles, we consider the following 
normalized Gaussian trial wave function for fermion component $j$ of
  (\ref{m})
\begin{equation}\label{vx}
\varphi_{vj}= \sqrt{\frac{1}{R_j(\tau)\sqrt \pi}} \exp\biggr[
-\frac{y^2}{2R_j^2(\tau)}
\biggr], \quad j=1,2. 
\end{equation}
Using essentially the Lagrangian density (\ref{yy}) in this case we 
obtain the following effective Lagrangian 
\begin{eqnarray}
L=-\frac{1}{\sqrt \pi}\frac{N_{jk}N_j}{\sqrt{R_1^2+R_2^2}}
+\sum_{j=1}^2\frac{N_j}{2}\biggr(\frac{6\sqrt 3}{5 \sqrt
5\pi^{1/3}}\frac{N_{jj}}{R_j^{2/3}} +\frac{1}{R_j^2}
\biggr),\nonumber   
\end{eqnarray}
with $j\ne k =1,2$.
The variational Lagrange equations (\ref{la}) for $R_1$ and $R_2$ now
become
\begin{eqnarray}\label{cp}
\frac{4}{R_j^3}-\frac{4N_{jk}}{\sqrt \pi}\frac{R_j}{(R_1^2+R_2^ 2)^{3/2}}
+\frac{8\sqrt 3}{5 \sqrt 5\pi ^{1/3}}
\frac{N_{jj}}{R_j^{5/3}}=0.
\end{eqnarray}
Equations (\ref{cp}) can be solved for variational widths $R_j$ and
consequently the variational profile of the wave functions obtained 
from   (\ref{vx}). When $N_1=N_2$, in   (\ref{cp})  $N_{jj}= \beta$,  
$N_{jk}=N_{kj}=\gamma$ and $R_1=R_2=R$; and in this case 
  (\ref{cp}) yields the same 
variational widths as from  the result obtained in the 
symmetric case in
section 4.1 given by   (\ref{min}); e. g.,  
\begin{equation}\label{sim}
\frac{4}{R^3}-\frac{\gamma}{R^2}\sqrt 
{\frac{2}{\pi}}+\frac{1}{R^{5/3}}\frac{8 \beta\sqrt 3}{5 
\pi^{1/3}\sqrt 5}=0.
\end{equation}

 \section{Numerical Results}

We solve   
  (\ref{m})  for bright  solitons
numerically using a time-iteration
method based on the Crank-Nicholson discretization scheme
 \cite{sk1}. We discretize the coupled partial
differential equations  (\ref{m})    
using time step $0.0002$ and space step $0.015$ in the domain 
$-8<y<8$. The second derivative in $y$ is discretized by a three-point
finite-difference rule and the first derivative in $\tau$  by a two-point
finite-difference rule.  We perform  a time evolution of    (\ref{m}) 
 introducing an harmonic oscillator potential $y^2$ in it 
and setting the nonlinear terms to
zero, and  
starting with the eigenfunction of the  harmonic
oscillator problem: $\varphi_i(y,\tau) 
=
\pi^{-1/4}\exp(-y^2/2)\exp(-i\tau).$  
The extra 
harmonic oscillator potential, set equal to zero 
in the end,
only aids in starting the time evolution
with an exact analytic form. With this initial solution we perform time
evolution of    (\ref{m}).     
During the time evolution the nonlinear
terms are  switched on  slowly and  the harmonic oscillator potential is
switched off slowly and 
the time evolution  continued to obtain the final converged
solutions.  
In addition to solving the coupled equations 
(\ref{m}), we also solved the single equation (\ref{o}) in the symmetric
case: $N_1=N_2$.
For given values of $N_1$ and $N_2$, solitons can be obtained for
$|a_{12}|$ above a certain value. 
In the coupled-channel case solitons are easily obtained for a
smaller $|a_{12}|$ when $N_1$ and $N_2$ are not very different from each
other.

\begin{figure}%[!ht]
 
\begin{center}
\includegraphics[width=.78\linewidth]{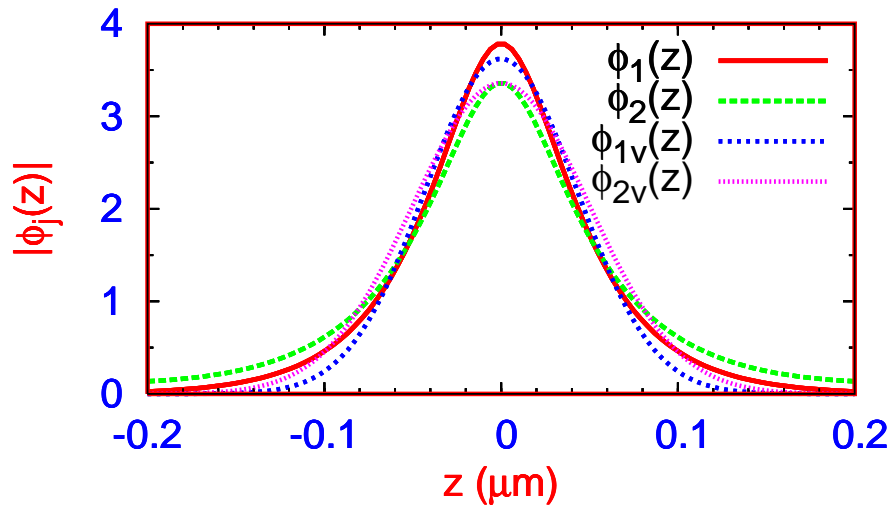}
\end{center}

\caption{The solitons $|\phi_j(z)|$ of  (\ref{m})
  vs. $z$ for
 $N_1=40$, $N_2=60$, $a_{12}=-300 $ nm, while   
$N_{11} \approx 149 $, $N_{12}=  216$, $N_{21}=  144$, and
$N_{22}\approx 195$. 
The corresponding variational solutions $\phi_{jv}(z)$ calculated from
the coupled equations  (\ref{cp}) 
  are also shown. 
} \end{figure}

In our numerical study we take $l=1$ $\mu$m and 
consider a DFFM consisting of two electronic states of 
 $^{40}$K  atoms. This  corresponds to a radial
trap of frequency $\omega= \hbar/(l^2m) \approx 2\pi \times 83$ Hz. 
Consequently, the  unit of
time is  $2/\omega \approx 4$ ms. 
For another fermionic atom the $\omega$ value 
gets changed accordingly for $l=1$ $\mu$m.

\subsection{Single Soliton}

From a solution of coupled  equations  (\ref{m}) we find that these
equations 
permit solitonic solutions provided that $|a_{12}|$ is larger than a
certain threshold value consistent with the analysis of section 3. 
First we solve    (\ref{m}) for   
$N_1=40  , N_2= 60$, 
and $a_{12}=-300$ nm. In this case no solitons are allowed for
$a_{12}=-290$ nm. The solitonic solution suddenly appears as  $|a_{12}|$
increases past $290$ nm. 
 The solitons in this case are shown in figure   1, where we also
plot the variational solutions of coupled equations  (\ref{cp}). 
In this case the nonlinearity parameters are 
 $N_{11} \approx 149 $, $N_{12}=  216$,
$N_{21}=  144$, and
$N_{22}\approx 195$.
Substituting these
values of the nonlinearities in    (\ref{cp}) and solving  we obtain
the variational
widths $R_1 \approx 0.043$ and  $R_2 \approx 0.050$. 
From   (\ref{v}) and (\ref{con}) we then obtain the   
variational
soliton profiles 
$|\phi_{v1}(z)| \approx 3.62\exp(-270z^2)$ and 
$|\phi_{v2}(z)| \approx 3.36\exp(-200z^2)$. 
From figure  1 we find that the
variational results agree 
well with the numerical solutions of the coupled equations. 
In this case
we also found the
variational result in the symmetric case ($N_1= N_2= 50$)
using   (\ref{sim}),  
which lies close to  the above two variational solutions.

\begin{figure}%[!ht]
 
\begin{center}
\includegraphics[width=.58\linewidth]{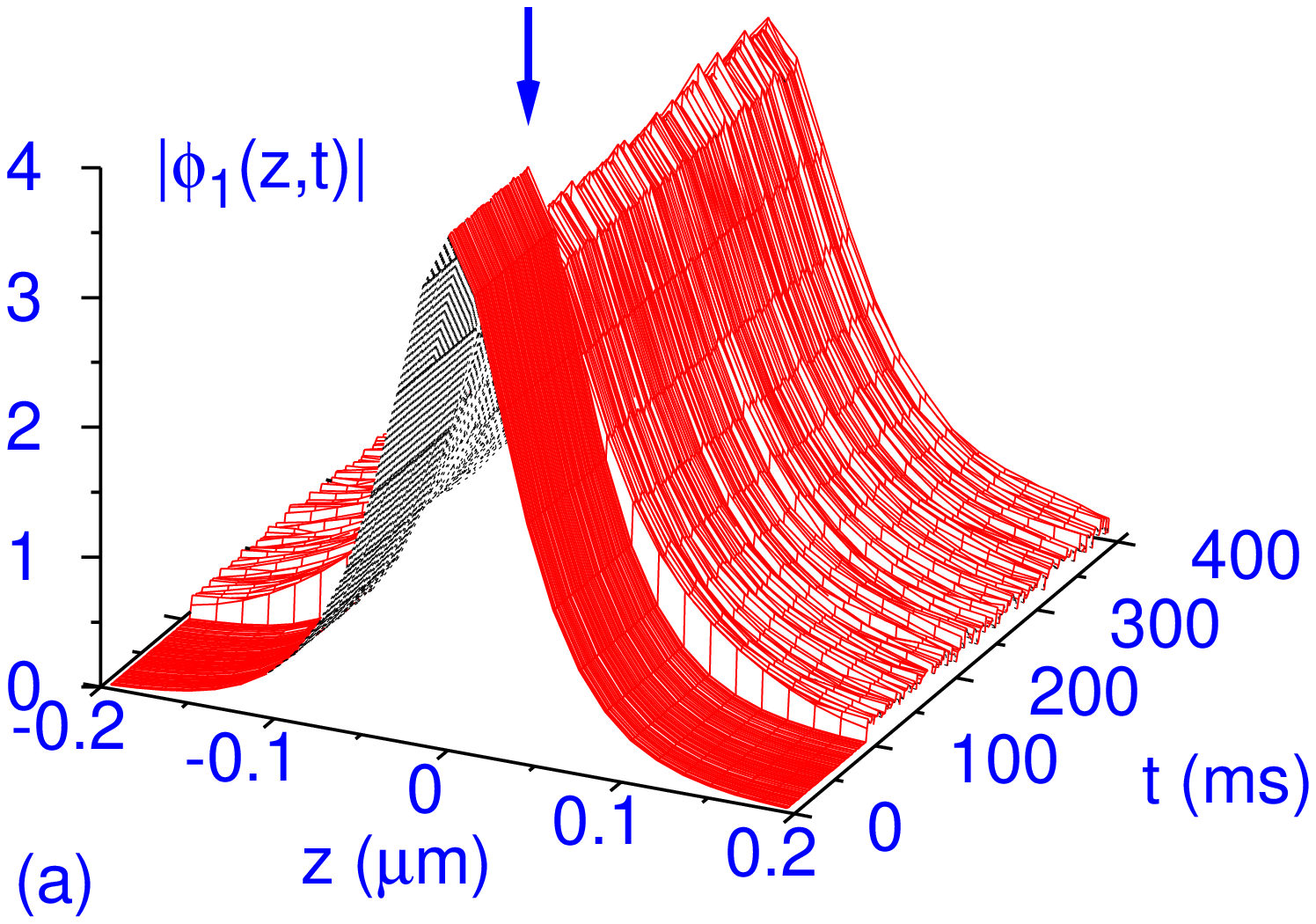}
\includegraphics[width=.58\linewidth]{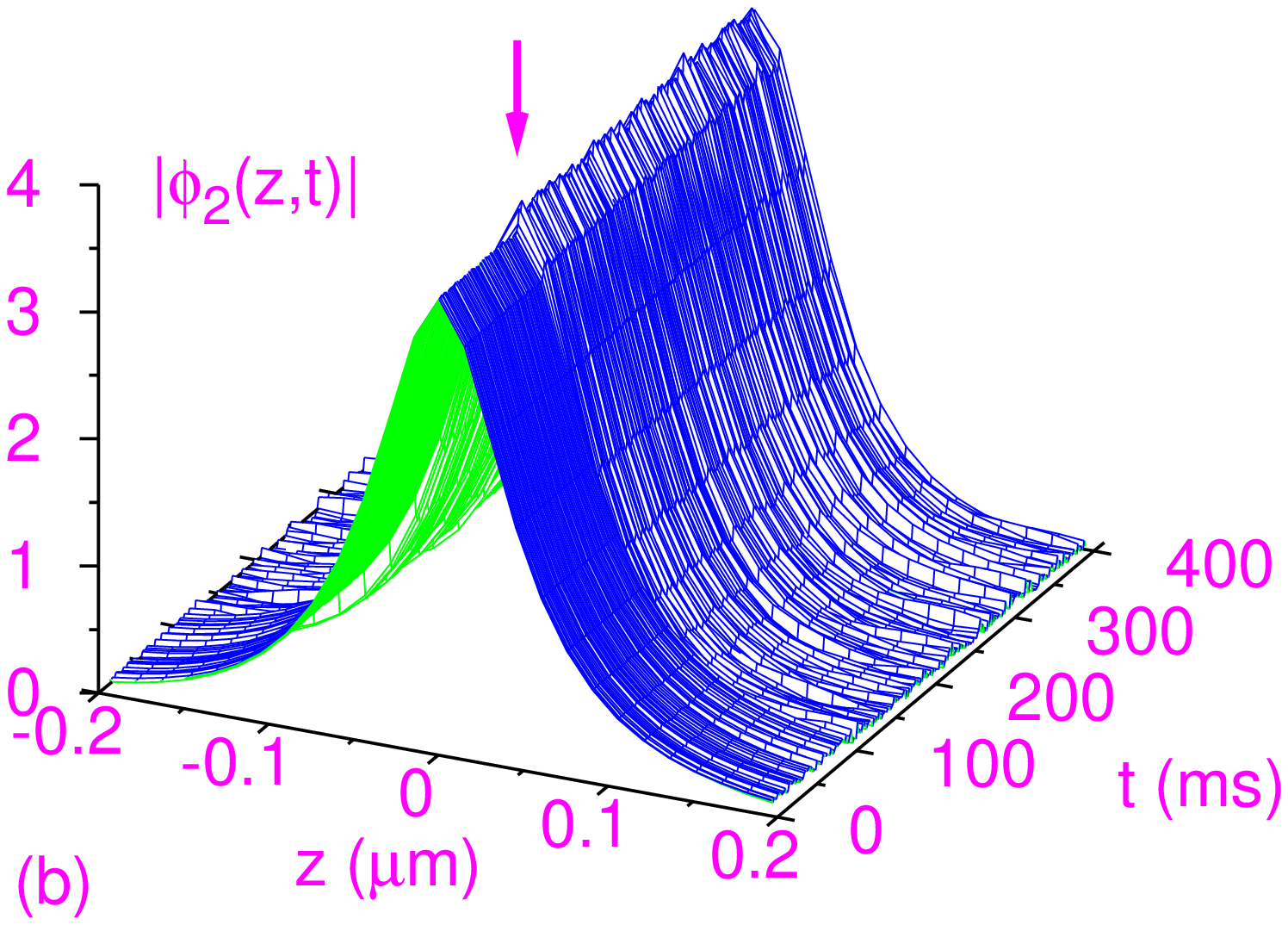}
\end{center}

\caption{ The propagation of fermionic
solitons (a)  $|\phi_1(z,t)|$ and (b)  $|\phi_2(z,t)|$ of figure  1 
  vs. $z$  and $t$ for $N_1=40$ and $N_2=60$. At
$t=100$ ms (marked by arrows) the bright solitons are set into small
breathing oscillation by suddenly  changing the fermion numbers to 
 $N_1=60$ and $N_2=40$.}
\end{figure}

To test the robustness of these solitons we
inflicted different perturbations on  them and studied the resultant
dynamics numerically. 
First, after the formation of the solitons of figure   1 
with $N_1=40$ and $N_2=60$ we
suddenly changed the particle numbers to  $N_1=60$ and $N_2=40$
at time $t= 100$ ms.  
This corresponds to a sudden
change
of nonlinearities from $N_{11}\approx 149 $, $N_{12}=
216$, $N_{21}=
144$, and $N_{22} \approx 195$ to
$N_{11}\approx 195 $, $N_{12}=  144$, $N_{21}= 216$, and
$N_{22} \approx 149$.
The resultant dynamics is 
shown in  figures   2a and 2b. 
Due to
the sudden change in nonlinearities, the fermionic bright
solitons are set into stable non-periodic small-amplitude
breathing oscillation.   This demonstrates the robustness of the solitons.

After the formation of the solitons  of figure 1 with $N_1=40$ and
$N_2=60$
we
suddenly changed the interspecies scattering length $a_{12}$ from 
$-300$
nm
to $-330$ nm
at time $t= 100$ ms.  
This can be realized by manipulating a background
magnetic field near a fermion-fermion Feshbach resonance \cite{fesh}
This corresponds to a sudden
change
of nonlinearities from $N_{11}\approx 149 $, $N_{12}=
216$, $N_{21}=
144$, and $N_{22} \approx 195$ to
$N_{11}\approx 149 $, $N_{12}\approx  238$, $N_{21}\approx 158$, and
$N_{22} \approx 195$. 
Due to
the sudden change in nonlinearities, the fermionic bright
solitons are set into stable non-periodic small-amplitude
breathing oscillation. Instead of plotting the soliton profile in this
case we plot the root mean square (rms) size of the solitons 
in figure  3 which demonstrates  the stable nonperiodic breathing
oscillation.  
We also (i) gave  a small displacement between the
centers of these solitons and (ii) suddenly changed  
$\phi_1 \to
1.1 \times \phi_1$ and  $\phi_2 \to
1.1 \times \phi_2$. In both cases 
after oscillation and dissipation the solitons 
continue stable propagation
 which shows their robust nature.

\begin{figure}%[!ht]
 
\begin{center}
\includegraphics[width=.77\linewidth]{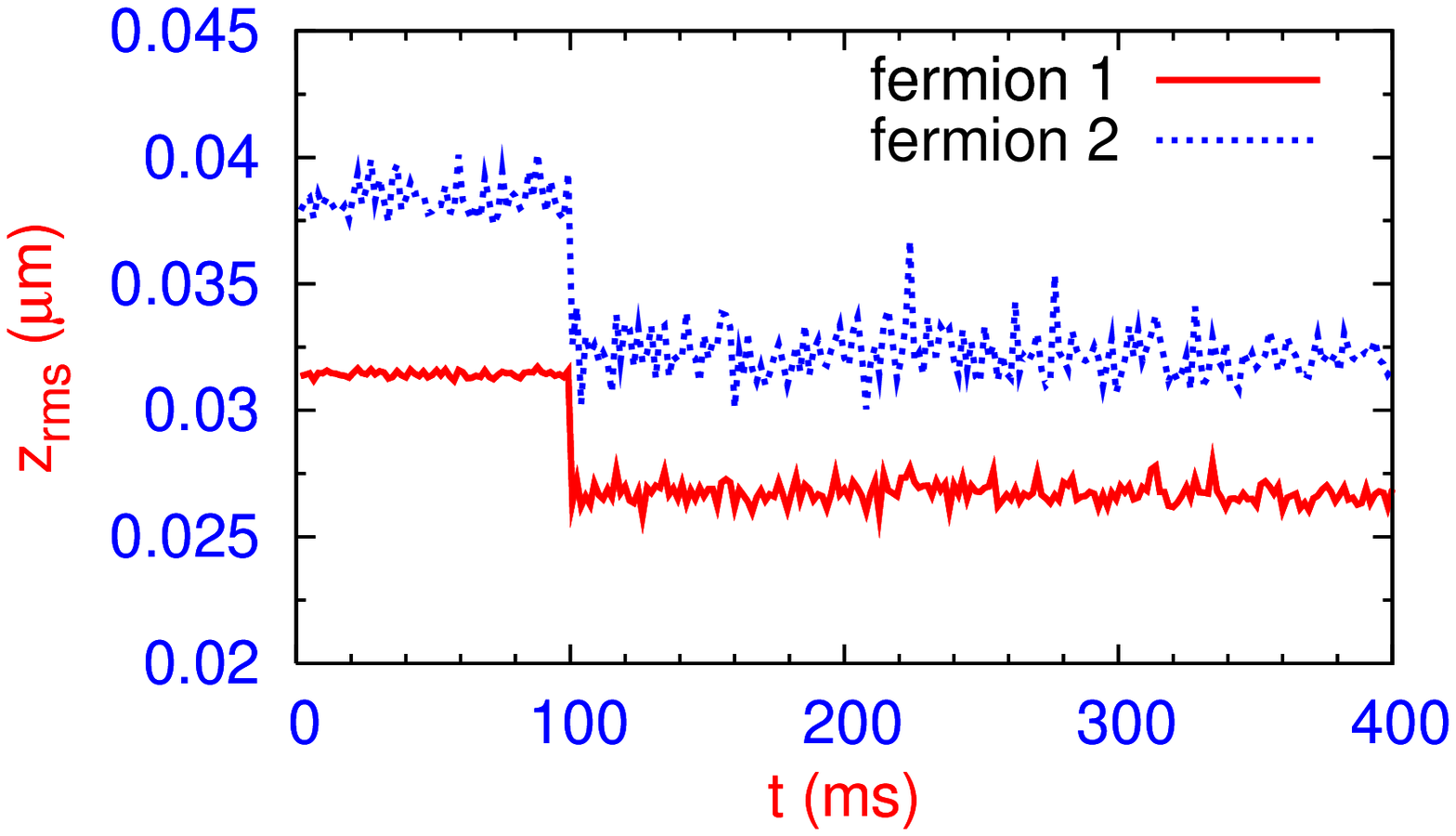}
\end{center}

\caption{  The rms sizes 
of the solitons   $|\phi_1(z,t)|$
and   $|\phi_2(z,t)|$ 
of figure  1
  vs. $t$.
At $t=100$ ms   the 
solitons of figure   1
are set into small
breathing oscillation by suddenly  changing  
$a_{12}$ from $-300$ nm  to $-330$ nm.   } \end{figure}

\subsection{Soliton Train}

\begin{figure}%[!ht]
 
\begin{center}
\includegraphics[width=.58\linewidth]{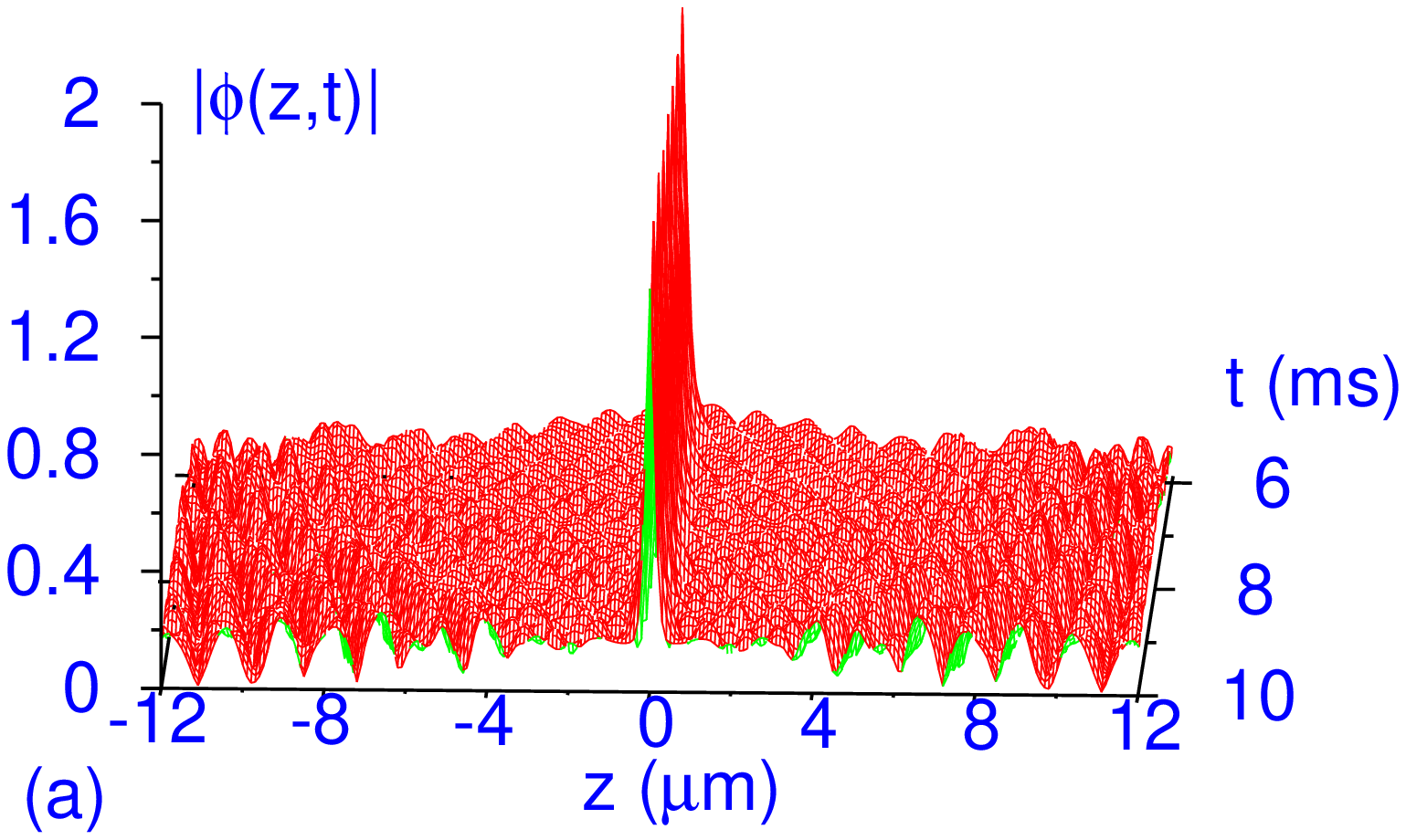}
\includegraphics[width=.58\linewidth]{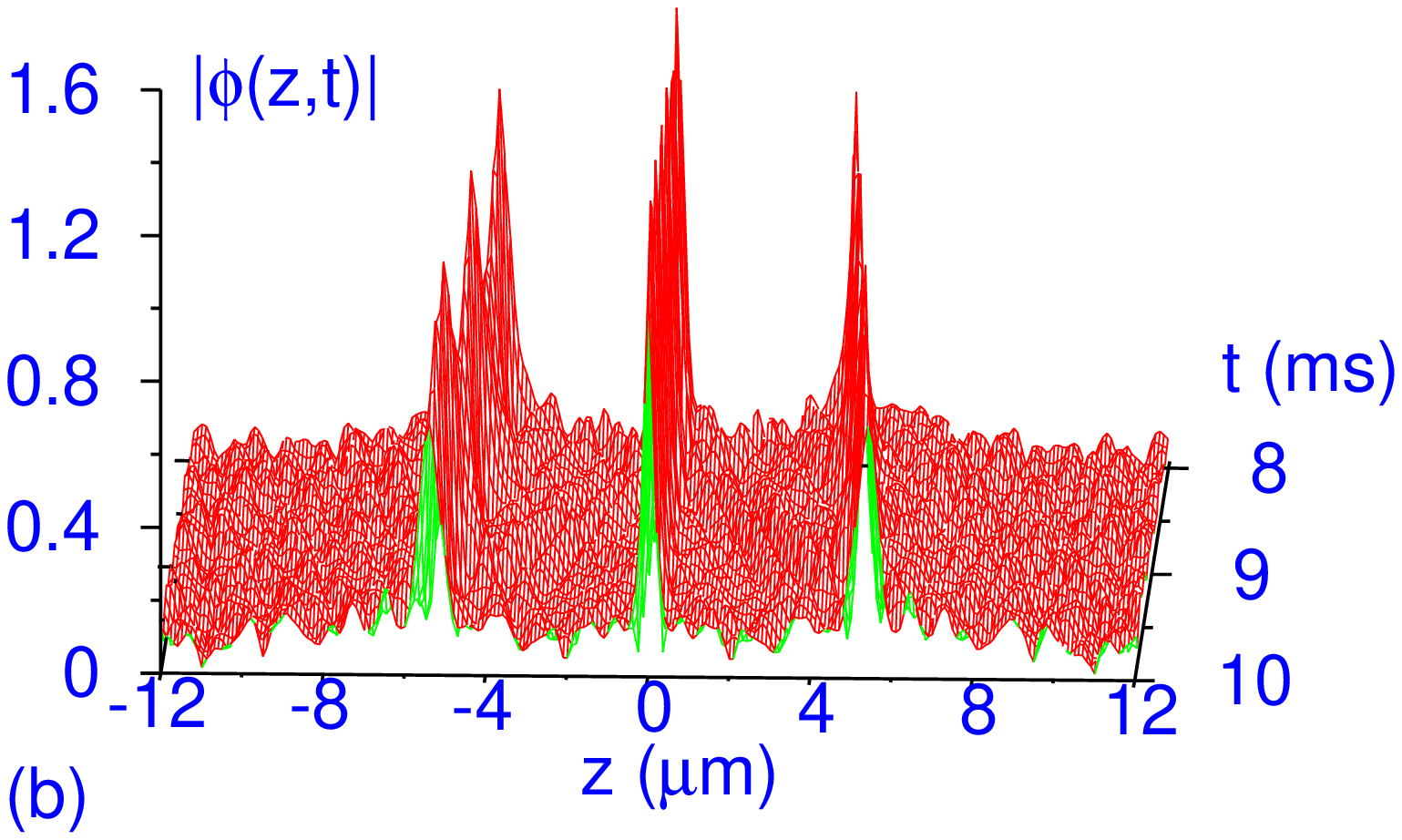}
\includegraphics[width=.58\linewidth]{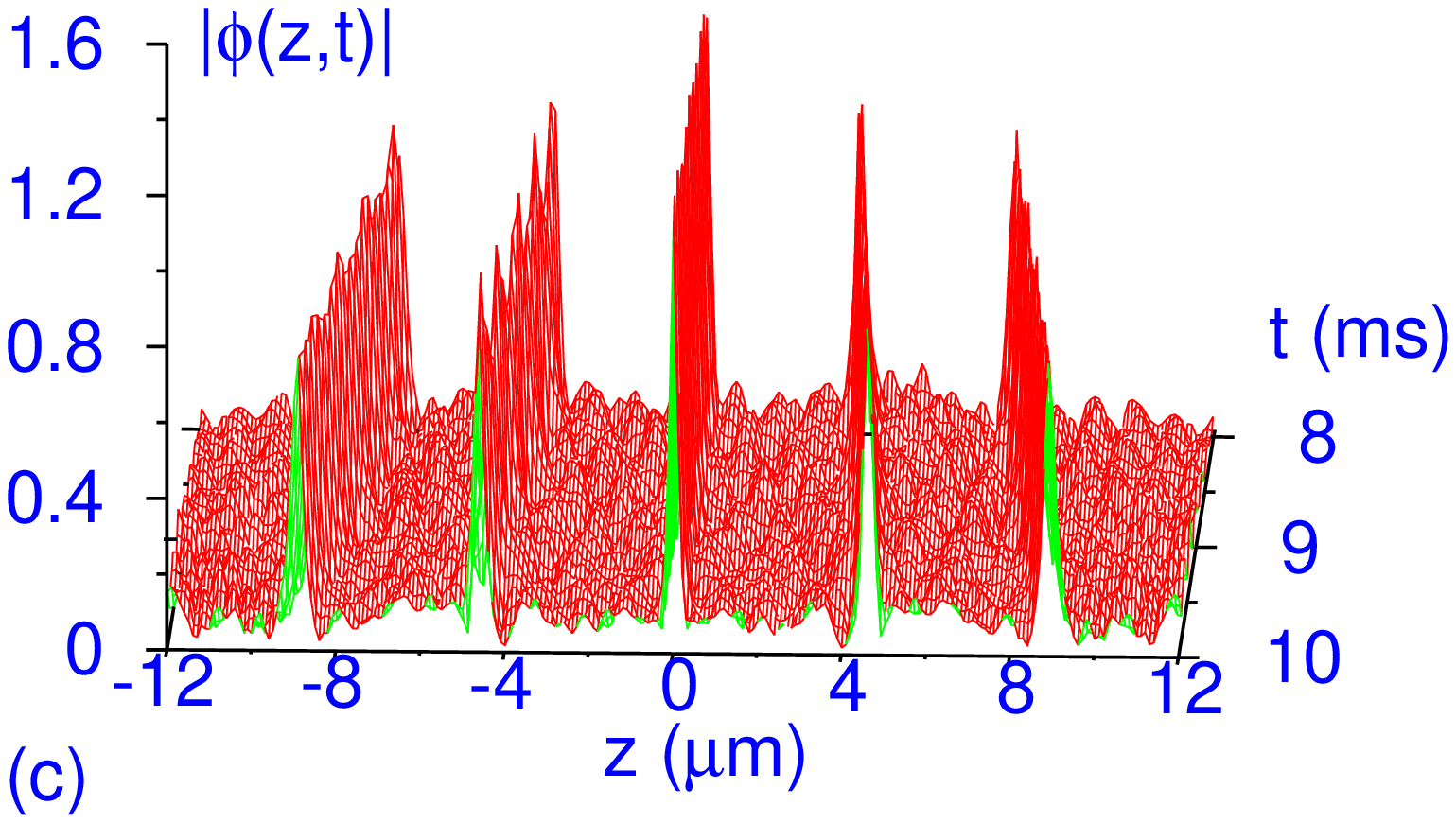}
\end{center}

\caption{Soliton trains of  one, three  and  five solitons
formed for $N_1=N_2=40$
upon removing the harmonic trap $y^2$ and 
jumping the nonlinearities at 
$t=0$ from $N_{jj}\approx 149, N_{jk}\approx 144,
k\ne j=1,2$ to 
(a)  $N_{jj}\approx 149, N_{jk}\approx 192,$
(b)  $N_{jj}\approx 149, N_{jk}\approx 274,
$ and to (c) $N_{jj}\approx 149, N_{jk} \approx 360,
$ respectively,  corresponding to a jump in scattering length 
$a_{12}$ from $-300$ nm to  $-400$ nm,  $-570$ nm, and $-750$ nm. 
For this purpose we solved the coupled equations (\ref{m}).} 
\end{figure}

During the time evolution of   (\ref{m}) 
if the 
nonlinearities  are changed 
by a small amount or changed slowly, usually one gets a single stable
soliton when the final nonlinearities are appropriate. 
However,
if the interspecies attraction is increased suddenly by a large 
amount by
jumping $a_{12}$ from a positive (repulsive) value to a large negative
(attractive) value, a
soliton train is formed  as in the experiment with BEC \cite{exdks1}
because of 
modulational instability \cite{tai}. 

To illustrate the formation of soliton train 
in a fermion-fermion mixture in numerical
simulation, 
we consider the solution of 
  (\ref{m}) for $N_1=N_2=40$ and $a_{12}=-300$ nm 
with an added axial harmonic trap $y^2$ 
corresponding to nonlinearities 
$N_{jj} \approx 149$ and $N_{jk}\approx 144, 
j\ne k=1,2$. 
After the formation of the solitons in the axial trap 
we
suddenly jump the scattering length to $-400$ nm, 
corresponding
to
off-diagonal nonlinearities $N_{jk}=  192$, 
 and also switch off the harmonic trap at time $t=0$. 
Although the initial value of scattering length ($a_{12}=-300$ nm)
in this case corresponds to an interatomic attraction, because of strong
Pauli-blocking repulsion the
effective fermion-fermion interaction is repulsive in this case. However,
the final
value of
scattering length ($a_{12}=-400$ nm)  corresponds to a stronger
interatomic
attraction which can overcome the Pauli
 repulsion
so that
an effective fermion-fermion attraction emerges in this case
which  might allow the formation of
soliton(s). In our numerical simulation we find that this is indeed
the case.  Upon  a jump of the scattering length to $a_{12}=-400$ nm,
we find that  after
some initial noise and dissipation
the
time evolution of   (\ref{m}) generates a single stable 
bright
soliton
as shown in figure  4a.

However, more solitons in the form of a soliton train can be 
formed  for a larger jump in the scattering length.
In figure   4b, from the same initial state in figure  4a we consider
a
larger jump of 
$a_{12}$ to $-570 $ nm  corresponding
to 
off-diagonal nonlinearities 
$N_{jk} \approx 274$. 
The final
value of
scattering length ($a_{12}=-570$ nm) in  this case  
corresponds to a
stronger
interatomic
attraction than considered in figure  4a 
which  might allow the formation of
soliton trains. This is verified in numerical simulation. 
Upon  a jump of the scattering length to $a_{12}=-570$ nm,
from  $a_{12}=-300$ nm,
we find that  after
some initial noise and dissipation 
the
time evolution of    (\ref{m}) generates three slowly receding  
bright 
solitons 
as shown in figure  4b. 
More solitons can be generated when the jump in the scattering length
$a_{12}$ or off-diagonal 
nonlinearities is
larger.

In figure 4c we show the generation of five  receding
solitons  of each component     
upon a
sudden jump of the scattering length 
$a_{12}$ to $-750$ nm from the
initial state of 
figure  4a. This
corresponds to a jump of 
the off-diagonal nonlinearities to
$N_{jk} \approx 360$. 
The formation of soliton trains from a stable initial 
state is due to
modulational instability \cite{1}.
The sudden jump in the off-diagonal nonlinearities could be
effected by a jump in the interspecies scattering length 
$a_{12}$ obtained
by
manipulating a background magnetic field near a 
 Feshbach resonance \cite{fesh}.

\section{Summary}

We used a coupled mean-field-hydrodynamic model for a DFFM
to study the formation of bright solitons and soliton trains in a
quasi-one-dimensional geometry by numerical and variational methods. We
find that an attractive interspecies interaction can overcome the Pauli
repulsion and form fermionic bright solitons in a DFFM. This is
similar to the formation of bright solitons in a coupled boson-boson
\cite{perez}
and
boson-fermion \cite{fbs1,fbs2} mixtures supported by interspecies
interaction.
We show by a linear stability analysis that when the interspecies
attraction is larger than a threshold
value, bright solitons can be formed due to modulational instability of a
constant-amplitude solution of the nonlinear equations describing the
DFFM.

The
stability of these solitons is demonstrated numerically through
their sustained breathing oscillation initiated by a sudden small
perturbation. We also illustrate the creation of soliton trains upon a
sudden large jump in interspecies attraction by manipulating a background
magnetic field near a Feshbach resonance \cite{fesh}
resulting in a sudden jump in
the 
off-diagonal nonlinearities.  
This jump transforms an effectively repulsive DFFM 
into an effectively attractive one, responsible for the formation of a
soliton train due to modulational instability. 
Bright solitons and
soliton trains have been created experimentally in attractive BECs in the
presence of a radial trap only without any axial trap \cite{exdks1}
in a similar fashion by transforming a repulsive BEC into an
attractive one.  In
view of this, fermionic bright solitons and trains could be created in
laboratory in a DFFM in a quasi-one-dimensional configuration.  

Here we
used a set of mean-field equations for the DFFM.  A proper treatment of a
DFG or DFFM should be done using a fully antisymmetrized many-body Slater
determinant wave function \cite{yyy1,fbs1,Mario} as in the case of 
scattering
involving many electrons \cite{ps}. However, in view of the success of a
fermionic mean-field model 
in studies of collapse \cite{ska}, bright \cite{fbs2} 
and dark  \cite{fds} soliton 
in a DBFM and of mixing and demixing in a DFFM \cite{mix},
we
do
not believe that the present study on bright solitons in a DFFM to be so
peculiar as to have no general validity.

%\ack
 
\ack{
The work is 
supported in part by the CNPq and FAPESP
of Brazil.}

\end{document}